\documentclass[fleqn,10pt]{SelfArx} 

\definecolor{color1}{RGB}{0,0,90} 
\definecolor{color2}{RGB}{0,20,20} 

\usepackage{hyperref} 
\hypersetup{hidelinks,colorlinks,breaklinks=true,urlcolor=color2,citecolor=color1,linkcolor=color1,bookmarksopen=false,pdftitle={Title},pdfauthor={Author},urlcolor=blue}

\usepackage{graphicx}
\usepackage[square]{natbib}
\usepackage{amsmath}
\usepackage{multirow}
\usepackage{subcaption}
\usepackage{pdflscape}
\usepackage{epstopdf}
\usepackage[final]{pdfpages}

\usepackage{graphicx}
\usepackage{natbib}
\usepackage{amsmath}
\usepackage{amsfonts}
\usepackage{multirow}
\usepackage{tikz}
\usetikzlibrary{calc}

\usepackage{mathtools,amssymb,lipsum}

\usepackage{cuted}
\usepackage{boondox-calo}

\usepackage{calligra}

\DeclareMathAlphabet{\mathcalligra}{T1}{calligra}{m}{n} \DeclareFontShape{T1}{calligra}{m}{n}{<->s*[2.2]callig15}{}

\JournalInfo{Published in Applied Physics Letters, 127, 012402 (2025)} 
\Archive{\href{https://doi.org/10.1063/5.0267494}{DOI: 10.1063/5.0267494}} 

\PaperTitle{Magnetic levitation at low rotation frequencies using an on-axis magnetic field} 

\Authors{Joachim Marco Hermansen$^{1}$, Frederik Laust Durhuus$^{2}$, Rasmus Bjørk$^{1,*}$} 
\affiliation{$^1$\textit{Department of Energy Conversion and Storage, Technical University of Denmark - DTU, DK-2800 Kgs. Lyngby, Denmark}} 
\affiliation{$^2$\textit{Department of Physics, Technical University of Denmark - DTU, DK-2800 Kgs. Lyngby, Denmark}} 
\affiliation{*\textbf{Corresponding author}: rabj@dtu.dk} 

\newcommand{\keywordname}{Keywords} 

\Abstract{The Ucar effect is a simple yet astonishing phenomenon where a permanent magnet can be levitated by placing it in the vicinity of another permanent magnet that rotates sufficiently fast. The few previous works on this type of magnetic levitation all required magnets rotating on the order of 200 Hz. Here we investigate the influence of applying a static magnetic field on the rotation axis and show that this can lower the needed rotation frequency to below 50 Hz. 
We explain this by a detailed analysis of the force producing levitation, which is a superposition of a repelling force caused by the off-axis (rotating) magnetic field and an attractive force due to the on-axis field. We study this force and resulting levitation experimentally, analytically and numerically for three different rotor magnet configurations, showing that trends in the levitation distance and frequency range can be accurately predicted from both the numerical and analytical models.}

\begin{document}

\flushbottom 

\maketitle 

\thispagestyle{empty} 

Magnetic levitation, well known from science fiction literature, is a technology that is becoming increasingly mature. Although Earnshaw's theorem prevents stable levitation of static ferromagnetic systems, levitation can be achieved by active stabilization \cite{de_boeij_contactless_2009,jansen_magnetically_2008,gupta_applications_2011}, with an induced current, as is known from Maglev trains \cite{hyung-woo_lee_review_2006}, with time varying magnetic fields as in the magnetic Paul trap \cite{sackett_magnetic_1993,perdriat_planar_2022,janse_characterization_2024} or by spin stabilised levitation using the gyroscopic effect \cite{michaelis_stability_2015,michaelis_horizontal_2014}. 

\begin{figure}[!t]
\begin{center}
\includegraphics[width=.50\textwidth]{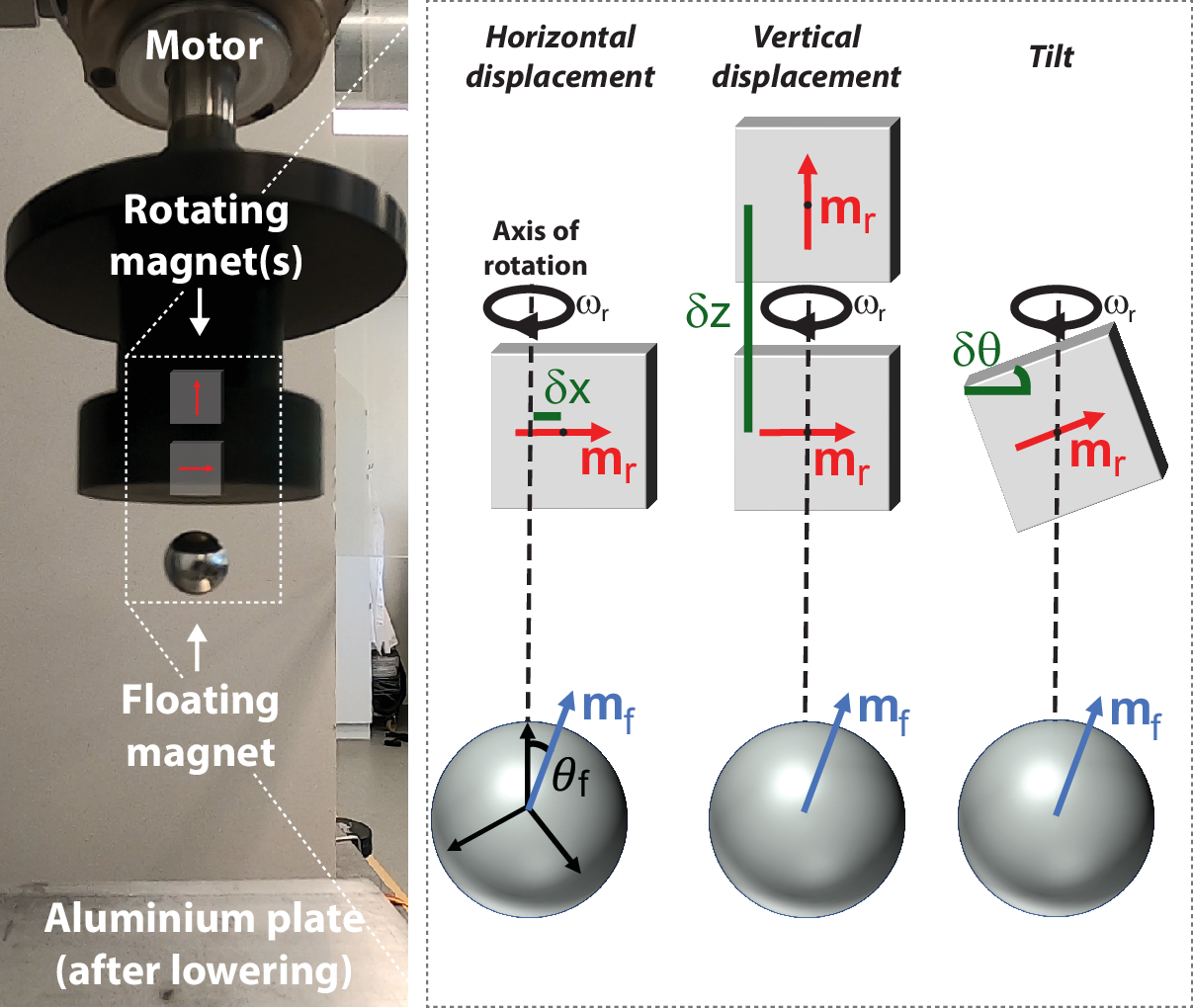}
\caption{The experimental setup, as seen from the camera recording the levitation. The floater magnet can clearly be seen to levitate. The rotating magnet(s) are embedded in a 3D-printed holder and cannot directly be seen. Therefore these have been indicated on the image. The three different rotor configurations investigated are shown on the right.}
\label{Fig.Exp_setup} 
\end{center}
\end{figure}

In the latter category a particularly interesting phenomenon - which we term the Ucar effect after its discoverer - was observed in 2021 by Ucar \cite{ucar_polarity_2021} and expanded on by us in Ref. \cite{Hermansen_2023}; using a simple setup with a rotating permanent magnet, another magnet can be made to levitate. In the most studied version a single magnet, termed the ``rotor'', is mounted on a rotating axis with its north and south poles perpendicular to the rotation axis and rotated on the order of $200\;\text{Hz}$. A second magnet, termed the ``floater'', is then placed near the rotor. Due to the magnetic torque, this is spun in motion and quite surprisingly an equilibrium of magnetic forces is automatically established, causing the floater to levitate a few centimeters below the rotor while precessing at the rotors rotation frequency. We illustrate the phenomenon in the image on the left side of Fig. \ref{Fig.Exp_setup}, which shows a floating magnet levitating below a rotating magnet. A video showing levitation is also available at e.g. Ref. \cite{YoutubeVideos}. The phenomenon can easily be reproduced with off-the-shelf components.

The Ucar effect occurs because the magnets are spinning. In the classical case where the rotor is static or rotating slowly, the floaters magnetic moment aligns with the rotors field so the interaction becomes attractive. If instead the magnets are spinning rapidly an inertial torque appears in the rotors rest frame, which balances the magnetostatic interaction such that the moments are kept at a position dependent equilibrium angle\cite{Hermansen_2023}.

As detailed below this angle scales with distance such that the magnetostatic repulsion is greater at short range relative to attraction, hence stable equilibrium points occur in mid-air just from rotational inertia and magnetostatic interaction. It was noted by Le Lay et al.\ \cite{le_lay_magnetic_2024} that the inertial torque can be decomposed into a gyroscopic component towards the axis of rotation and a centrifugal torque away from it, and unlike other spin stabilized levitation phenomena, the centrifugal component is the stabilizing part. 

As first mentioned by Ucar \cite{ucar_polarity_2021} and clarified by us \cite{Hermansen_2023} and Le Lay et al.\ \cite{le_lay_magnetic_2024} a constant magnetic field component along the rotation axis is crucial for levitation to occur.
This magnetic field component is denoted ``vertical field'' in Refs. \cite{le_lay_magnetic_2024,Hermansen_2023,ucar_polarity_2021}, however upon rotating the entire setup at an arbitrary angle to vertical, levitation can still occur hence ``on-axis magnetic field'' is more correct. The on-axis field provides the attractive component of the magnetic force needed to balance the repelling magnetic force caused by the rotating magnetic field. We showed in Ref. \cite{Hermansen_2023} that a mm scale imperfection in placement of the rotor magnet relative to the rotation axis is enough to produce levitation. However, while Ref. \cite{le_lay_magnetic_2024} studied equilibrium properties with a purposely tilted rotor magnet, the influence of the on-axis field on levitation stability is not known. 

In this work we investigate the effect of the on-axis field on levitation both experimentally, with analytical theory and using simulations. We show that the minimal rotor frequency decreases to about a third as the field strength of the on-axis magnetic field is changed. A nearby metal block can greatly dampen the floater magnets motion as eddy currents are induced in the block, which increases stability\cite{Hermansen_2023,ucar_polarity_2021,le_lay_magnetic_2024}. Here we show that by tuning the on-axis field, the levitation can be made to extend for very long times even in low-damping conditions.

We use the experimental setup described in Ref. \cite{Hermansen_2023}. In the setup a 3D printed plastic holder with the desired configuration of permanent magnets was mounted onto the shaft of a high speed motor (Vevor JST-JGF-F65A). The experimental setup is shown in Fig. \ref{Fig.Exp_setup}. 

We investigate three different permanent magnet configurations of the rotor, and for each vary the geometric parameter that changes the on-axis magnetic field. The configurations are as follows:
\begin{itemize}
    \item Horizontal displacement, where the rotor magnet is moved a distance $\delta{}x$ away from the rotation axis.
    \item Vertical displacement, where an additional rotating magnet with an on-axis magnetization is displaced a distance $\delta{}z$ along the rotation axis.
    \item Tilt, where the rotating magnet is tilted with an angle $\delta{}\theta$ relative to the rotation axis.
\end{itemize}
These three configurations are illustrated in Fig. \ref{Fig.Exp_setup} along with the geometrical parameter varied. The horizontal displacement parameter $\delta{}x$ was varied from 1 mm to 8 mm in steps of 1 mm. The vertical displacement parameter $\delta{}z$ was varied from 10 mm (the two magnets touching) to 20 mm in steps of 2 mm. The tilt parameter $\delta{}\theta$ was varied from $5^\circ$ to $45^\circ$ in steps of $5^\circ$. The above configurations all produce a combination of an on-axis static magnetic field and an off-axis rotating magnetic field as shown in Fig. 1 in the supplementary material. However, levitation was only possible for $\delta x = 1-4\: \mathrm{mm}$, $\delta z = 10-16 \: \mathrm{mm}$ and $\delta \theta = 5-20^\circ$. 
We note that for the vertical displacement experiments, the additional magnet rotates along with the rotor, but because it is vertically magnetized, this has no influence on the dynamics.

The permanent magnet(s) in the rotor are NdFeB-type cube magnets with a size of $10\times{}10\times{}10$ mm$^3$ and a remanent magnetization of 1.29-1.32 T. The magnets are glued into a 3D-printed holder. The position of the magnet when moved from the center axis is adjusted with pieces of non-magnetic steel, such that the center of mass remains on the rotation axis. The floater is a spherical NdFeB magnet with a diameter of 12.7 mm and a remanence identical to the rotor magnets.

During experiments the dynamical behavior of the floater magnet was record using a GoPro Hero 8 recording at 30 frames per second. Subsequently the recordings were post-processed with a SAM2 \cite{ravi2024sam2segmentimages} machine learning code to segment the video. All videos recorded are available at the data repository for this work \cite{Data_2025}.

In experiments, the rotor magnet is mounted on the motor and spun at the chosen speed. To dampen out initial vibrations of the floater a 30 mm thick aluminium plate of lateral dimensions $100 \times 240 \mathrm{mm}^2$ is positioned with its center 35 mm below the rotor magnet, except at extreme frequencies where the position was adjusted to 20 or 45 mm for high and low frequencies respectively.
Once the floater is placed beneath the rotor, the floater quickly finds its equilibrium position and rotational speed. As soon as this is reached, the aluminium plate is lowered to a distance of 90 mm or more from the center of the rotor magnet at a constant speed of 10 mm/s and is thus only relevant during the initial levitation of the floater magnet. The frequency was varied from the lowest frequency at which levitation was possible and in steps between 1-10 Hz until levitation was no longer possible.

To simulate the floaters motion, we use the same model, algorithm and parameters as in Ref. \cite{Hermansen_2023}, i.e. we time-step integrate the Newtonian equations of motion with gravity, magnetic dipole-dipole interactions and damping terms proportional to linear- and angular velocity. 

The dipole force, $\mathbf{F}_\text{dip},$ experienced by the floater magnet is given by
\begin{align}
    \mathbf{F}_\text{dip} &= \frac{3\mu_0}{4\pi} \frac{1}{r^4}[(\mathbf{m}_\text{f} \boldsymbol{\cdot} \hat{\mathbf{r}}) \mathbf{m}_\text{r} + (\mathbf{m}_\text{r} \boldsymbol{\cdot} \hat{\mathbf{r}}) \mathbf{m}_\text{f}
    \notag\\
    &\quad +(\mathbf{m}_\text{f} \boldsymbol{\cdot} \mathbf{m}_\text{r})\hat{\mathbf{r}}  -5(\mathbf{m}_\text{f}\boldsymbol{\cdot} \hat{\mathbf{r}})(\mathbf{m}_\text{r} \boldsymbol{\cdot} \hat{\mathbf{r}}) \hat{\mathbf{r}}], \label{eq:F_dip}
\end{align}
where $\mathbf{r}$ is the displacement from rotor to floater, $\mathbf{\hat{r}} = \mathbf{r}/r$ is normalized displacement, $\mathbf{m}$ is the magnetic moment and the $r$ and $f$-subscripts denote rotor and floater, respectively. We note that while the rotor magnets are cubes, they are far enough from the floater that they can be considered as dipoles \cite{Smith_2010,Bjoerk_2021,Bjoerk_2023}. The dipole magnetic field from a rotor magnet is given by
\begin{align}
   \mathbf{B}_\text{r} &= \frac{\mu_0}{4\pi r^3} \left[3(\mathbf{\hat{r}} \boldsymbol{\cdot} \mathbf{m}_\text{r}) \mathbf{\hat{r}} - \mathbf{m}_\text{r}\right]. \label{eq:B_dip}
\end{align}
We denote the components parallel and perpendicular to the rotation axis as $B_{\text{r}, z}$ and $B_{\text{r}, \perp}$ respectively.

By solving Newtons equations of motion for the floater to leading order, we previously found that the polar angle of the floater, $\theta_\text{f}$, shown on Fig. \ref{Fig.Exp_setup}, is given by \cite{Hermansen_2023}
\begin{align}
    \theta_\text{f} = \frac{m_\text{f} B_{\text{r}, \perp}}{I_\text{f} \omega_\text{r}^2 - m_\text{f} B_{\text{r}, z}} 
\end{align}
where $\omega_\text{r}$ is the angular velocity of the rotor and $I_\text{f}$ is the moment of inertia of the floater. 

In the on-axis direction, both repulsion and attraction are magnetostatic, with gravity producing a slight shift of the equilibrium position. In the off-axis direction, our simulations indicate that the floater performs a small circular motion, such that the radial magnetic force is precisely balanced by a centrifugal force. In the supplementary material, we show that the radius, $a$, of this small side mode motion can be derived from Eq. \ref{eq:F_dip} as;
\begin{align}
    \quad a=\frac{\mathcal{G}}{\mu_\text{f} \omega_r^2} \quad \text{where} \quad \mathcal{G} = \frac{3\mu_0 m_\text{f} m_\text{r}}{4\pi d^4},\label{eq:theta_f}
\end{align}
with the mass of the floater being $\mu_\text{f}$ and $d$ the equilibrium levitation distance along the rotation axis, i.e.\ $d=|\mathbf{r} \boldsymbol{\cdot} \boldsymbol{\hat{\omega}}_\text{r}|$. For the typical magnetic moments, mass and rotation speeds in this study, the value of $a$ is $0.1-0.3$ mm. Including this lateral displacement in our on-axis force calculations improves the agreement with experiments.

We find the equilibrium levitation distance $d$ for the case of horizontal displacement first. We consider a system in steady state, and at the exact moment when the magnetization of the floater magnet is in the $xz$-plane.
Then 
\begin{align}
    \mathbf{r} = \begin{pmatrix}
        -\delta{}x+a \\ 0 \\ -d
    \end{pmatrix},     
    \mathbf{m}_\text{f} = \begin{pmatrix}
        m_\text{f}\sin \theta_\text{f} \\ 0 \\ m_\text{f}\cos\theta_\text{f}
    \end{pmatrix},
    \mathbf{m}_\text{r} = \begin{pmatrix}
        m_\text{r} \\ 0 \\ 0
    \end{pmatrix} 
\end{align}

Inserting these in Eq. \eqref{eq:F_dip} and using Eqs. \eqref{eq:B_dip} and \eqref{eq:theta_f} we get
\begin{align}
    F_{z,\text{hor}} = \mathcal{G} \Biggr( 
    \left[5\frac{(\delta{}x-a)^2}{d^2}-1 \right] \sin \theta_\text{f} +4\left[\frac{\delta{}x-a}{d} \right] \cos \theta_\text{f} \Biggr) \label{Eq.FzHDSemi}
\end{align}

Assuming the polar angle to be small, which implies $I_\text{f} \omega_\text{r}^2\gg m_\text{f} B_{\text{r}, z}$ and thus $\theta_\text{f} \approx \frac{m_\text{f} B_{\text{r}, \perp}}{I_\text{f} \omega_\text{r}^2}$, we obtain
\begin{align}
    F_{z,\text{hor}} &= \mathcal{G}\left( 4\frac{\delta{}x-a}{d} 
     -\frac{m_\text{f} B_{\text{r}, \perp}}{I_\text{f} \omega_\text{r}^2}  \right)\label{Eq.FzHD}
\end{align}

\begin{figure}[!t]
\begin{center}
\includegraphics[width=.50\textwidth]{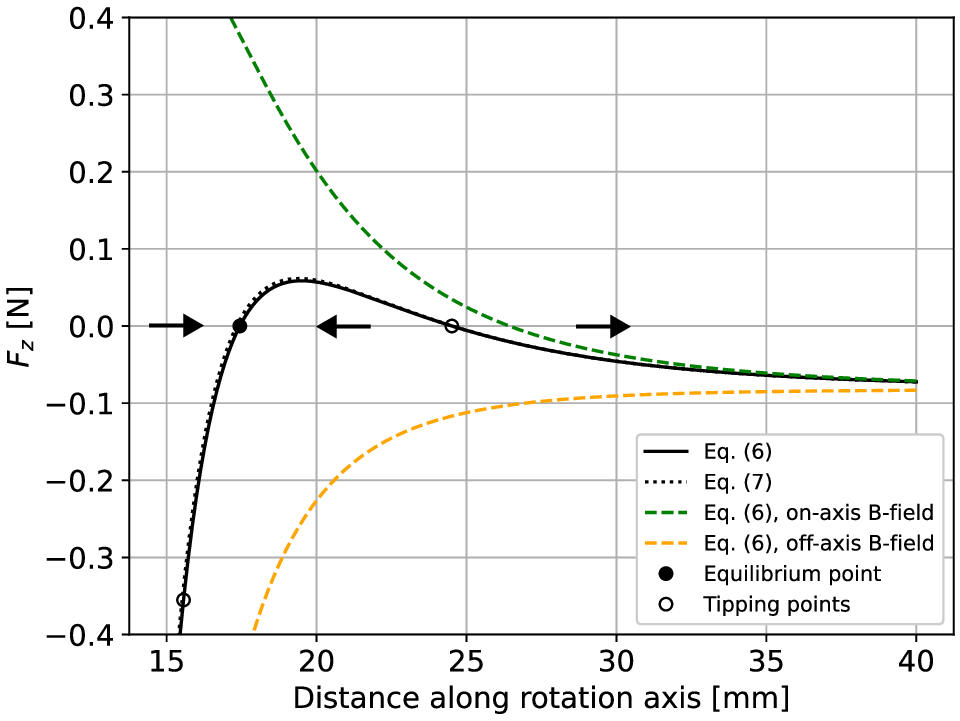}
\caption{The on-axis force and its associated magnetic field components for the case of a horizontal displacement of $\delta{}x=1$ mm with gravity, using Eqs. \eqref{Eq.FzHDSemi} or \eqref{Eq.FzHD}. The equilibrium point and the tippings points are also shown, along with arrows for the force direction.}
\label{Fig.Force_z_HD} 
\end{center}
\end{figure}

The force for a displacement of $\delta{}x = 1$ mm and including gravity, is shown in Fig. \ref{Fig.Force_z_HD}. The first term in Eq. \ref{Eq.FzHD} is caused by the on-axis field component, while the second term is caused by the off-axis (rotating) field component. Because $B_{\text{r},\perp} \sim d^{-3}$, the repulsive off-axis term is more strongly dependent on distance. At large distances the force, $F_z$, is equal to the value of gravity.  As seen from Fig. \ref{Fig.Force_z_HD} this implies a stable point where the vertical force is zero and around which the force is restoring, leading to stable levitation. There is a tipping point at large distance, where the force is no longer attracting, and a tipping point at short distance, since displacements to lower values than this will cause the floater to over-shoot the tipping point at long distance when oscillating. For this geometry, the angle $\theta_f$ changes from almost zero at $d=40$ mm to 9 degrees at the tipping point closest to the rotor. The other rotor configurations studied (see Fig. \ref{Fig.Exp_setup}), whose force expressions are derived subsequently produce qualitatively equivalent curves to Fig. \ref{Fig.Force_z_HD}, hence the arguments for stable levitation are identical.

For the vertical displacement configuration, there are two rotor magnets. The displacement vectors to the horizontal, $\mathbf{r}^\text{H}$, and vertical, $\mathbf{r}^\text{V}$, rotor magnets and the magnetic moments are given by
\begin{align}
    \mathbf{r}^\text{H} =\begin{pmatrix}
        a \\ 0 \\ -d
    \end{pmatrix}, \:
    \mathbf{r}^\text{V} =\begin{pmatrix}
        a \\ 0 \\ -(d+\delta{}z)
    \end{pmatrix}
\end{align}
and
\begin{align}
    \mathbf{m}_\text{f} = \begin{pmatrix}
        m_\text{f}\sin \theta_\text{f} \\ 0 \\ m_\text{f}\cos\theta_\text{f}
    \end{pmatrix},
    \mathbf{m}^\text{H}_\text{r} = \begin{pmatrix} 
        m^\text{H}_\text{r} \\ 0 \\ 0
    \end{pmatrix},
    \mathbf{m}^\text{V}_\text{r} = \begin{pmatrix} 
        0 \\ 0 \\ m^\text{V}_\text{r}
    \end{pmatrix} 
\end{align}
Inserting these in Eq. \eqref{eq:F_dip} and using Eqs. \eqref{eq:B_dip} and \eqref{eq:theta_f} we get for the $z-$component of the force
\begin{align}
    F_{z,\text{ver}} = \mathcal{G} \Biggr(&- \sin \theta_\text{f} -\frac{4a}{d} \cos \theta_\text{f} + 5   \frac{a^2}{d^2}\sin \theta_\text{f} 
       \nonumber \\ & - \frac{4ad^4 \sin \theta_\text{f} }{(d+\delta{}z)^5}    + \frac{2d^4 \cos \theta_\text{f}}{(d+\delta{}z)^4}     \Biggr)\label{eq:VD_forcezSemi}
\end{align}

Finally, for the tilt configuration where the rotor magnet is rotated an angle $\delta{}\theta$, the displacement vector, $\mathbf{r}$, and the magnetic moments are given by
\begin{align}
    \mathbf{r} = \begin{pmatrix}
        a \\ 0 \\ -d
    \end{pmatrix},
    \mathbf{m}_\text{f} = \begin{pmatrix}
        m_\text{f}\sin \theta_\text{f} \\ 0 \\ m_\text{f}\cos\theta_\text{f}
    \end{pmatrix},
    \mathbf{m}_\text{r}  = \begin{pmatrix}
        m_\text{r}\cos{\delta{}\theta} \\ 0 \\ m_\text{r}\sin{\delta{}\theta} 
    \end{pmatrix} 
\end{align}
 
Again inserting these in Eq. \eqref{eq:F_dip} and using Eqs. \eqref{eq:B_dip} and \eqref{eq:theta_f} we get for the $z-$component of the magnetic force
\begin{align}
    F_{z,\text{tilt}} &= \mathcal{G} \Bigg(2\cos \theta_\text{f} \sin \delta \theta - 4\frac{a}{d}\cos(\theta_\text{f} - \delta{}\theta) 
    \nonumber\\
    &\quad + \left[5\frac{a^2}{d^2} - 1\right] \cos \delta{}\theta \sin \theta_\text{f}  \Bigg)\label{eq:tilt_forceEqSemi}
\end{align}

Assuming that in radians $\delta{}\theta\ll 1$ and $\theta_\text{f}\ll 1$, and also $a \ll d$, we get
\begin{align}
    F_{z,\text{tilt}} &= \mathcal{G} \left( 2\delta{}\theta-\theta_\text{f} \right) \label{eq:tilt_forceEq}
\end{align}

By setting the force in Eqs. \eqref{Eq.FzHD} and \eqref{eq:tilt_forceEq} equal to zero analytical expressions for the levitation distance can be obtained. These expressions are given in Eqs. 17 and 36 in the supplementary material; the latter being identical to Eq. 12 of Ref. \cite{le_lay_magnetic_2024}. Alternatively Eqs. \eqref{Eq.FzHDSemi}, \eqref{eq:VD_forcezSemi} and \eqref{eq:tilt_forceEqSemi} can be numerically solved to find the equilibrium point, $F_z= \mu_\text{f} g$ where $g$ is gravitational acceleration, which gives the levitation distance. This can be compared with the full numerical model solving Newtons equation of motion previously mentioned, as well as the experimental results. Note that only the analytical solutions neglect gravity.

\begin{figure}[!t]
\begin{center}
\includegraphics[width=.50\textwidth]{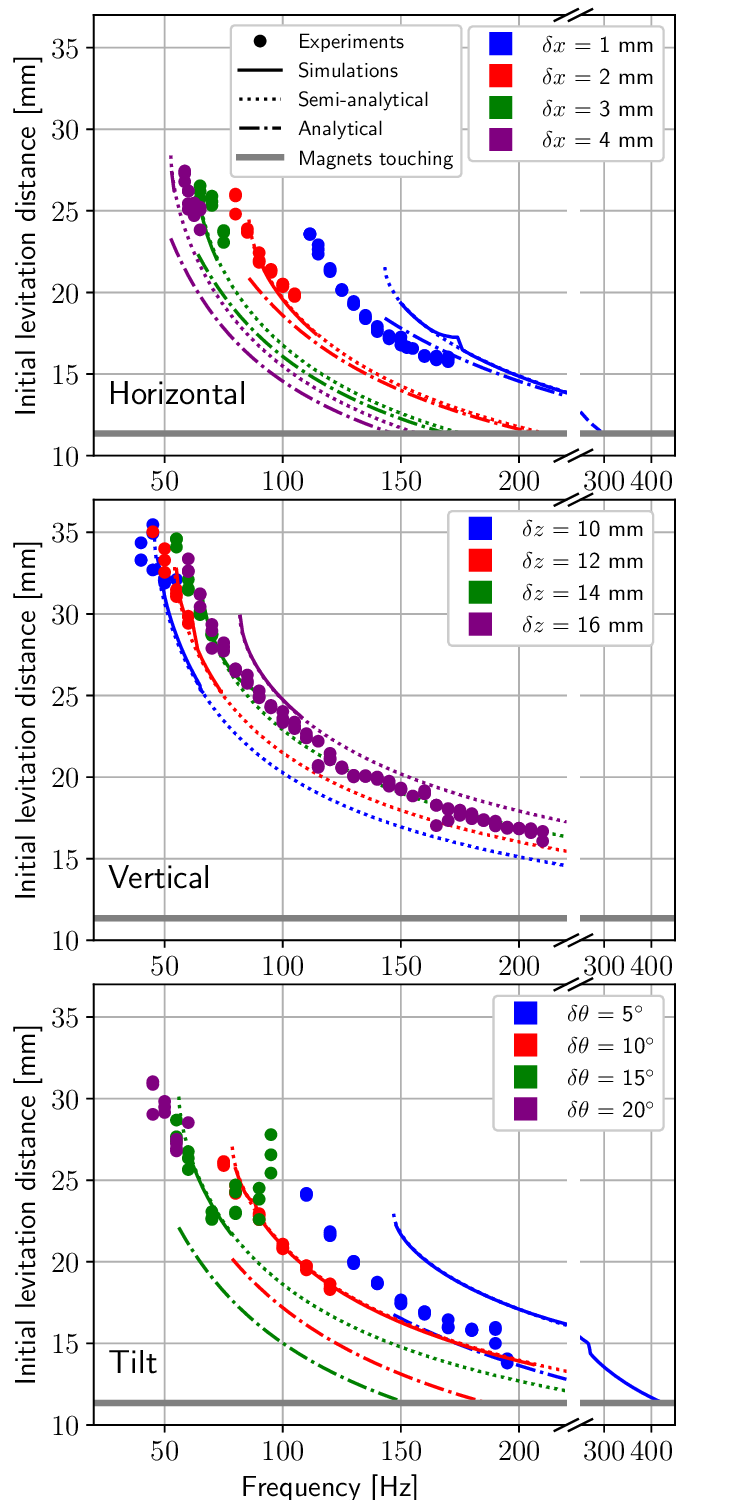}
\caption{The initial levitation distance for the three different rotor configurations, as function of rotor frequency. In each plot the geometric parameter in each setup is varied. Also shown is the levitation distance computed using the simulation framework, as well as the numerically-solved expressions given in Eqs. \eqref{Eq.FzHDSemi}, \eqref{eq:VD_forcezSemi} and  \eqref{eq:tilt_forceEqSemi} as well as the analytical solutions of Eqs. \eqref{Eq.FzHD} and \eqref{eq:tilt_forceEq}.}
\label{fig:Levitation_dist} 
\end{center}
\end{figure}

Shown in Fig. \ref{fig:Levitation_dist} is the initial levitation distance as a function of frequency for the three different rotor configurations for both the experiments, the simulation model and the above analytical expressions. As shown in Fig. 3 in the supplementary material, the levitation distance changes as function of time. What we show in Fig. \ref{fig:Levitation_dist} is the initial levitation distance after the aluminium plate is lowered. Each experiment was repeated three times. Of the tested parameters only those indicated in the legend allowed levitation, nor was levitation achieved for any frequencies outside the plotted ranges. We define levitation as the floater levitating for a minimum of 3 s before falling away from the rotor.

As can be seen from the figure, the frequency needed to achieve levitation is significantly reduced when the geometrical parameter is increased for both horizontal displacement and tilt, while the reverse tendency is true for vertical displacement. This is because for the former two, an increased geometrical parameter corresponds to an increasing on-axis field, while this is opposite for vertical displacement, as per Fig. 1 in the supplementary material. Additionally, we also observe an increasing initial levitation distance with lower frequency which was also observed in Ref. \cite{Hermansen_2023}. 

The simulations reproduce the experimental trends with respect to the change in frequency range as function of geometrical parameter and levitation distance vs.\ frequency. Simulations and semi-analytical solutions are in excellent agreement regarding levitation distance and we expect the deviation from experiments is largely due to the high parameter sensitivity; for instance sub-millimeter precision in $\delta x$ is required for precise comparison.  For horizontal displacement, the analytical solution also agrees with simulations at high frequencies where $\theta_\text{f} \sim \omega_\text{f}^{-2}$ is small, but at lower frequencies, the small angle approximation becomes imprecise. The tilt solution is less accurate throughout, because we also neglect a term linear in $a/d$, but both analytical expressions reproduce general trends despite neglecting gravity. The simulations also predict a far larger frequency range over which levitation is possible than the experiments show. This is most likely because the simulations do not display the destabilizing oscillations seen experimentally at high frequencies\cite{Hermansen_2023}, which might be related to eddy current effects.

\begin{figure}[!t]
\begin{center}
\includegraphics[width=.50\textwidth]{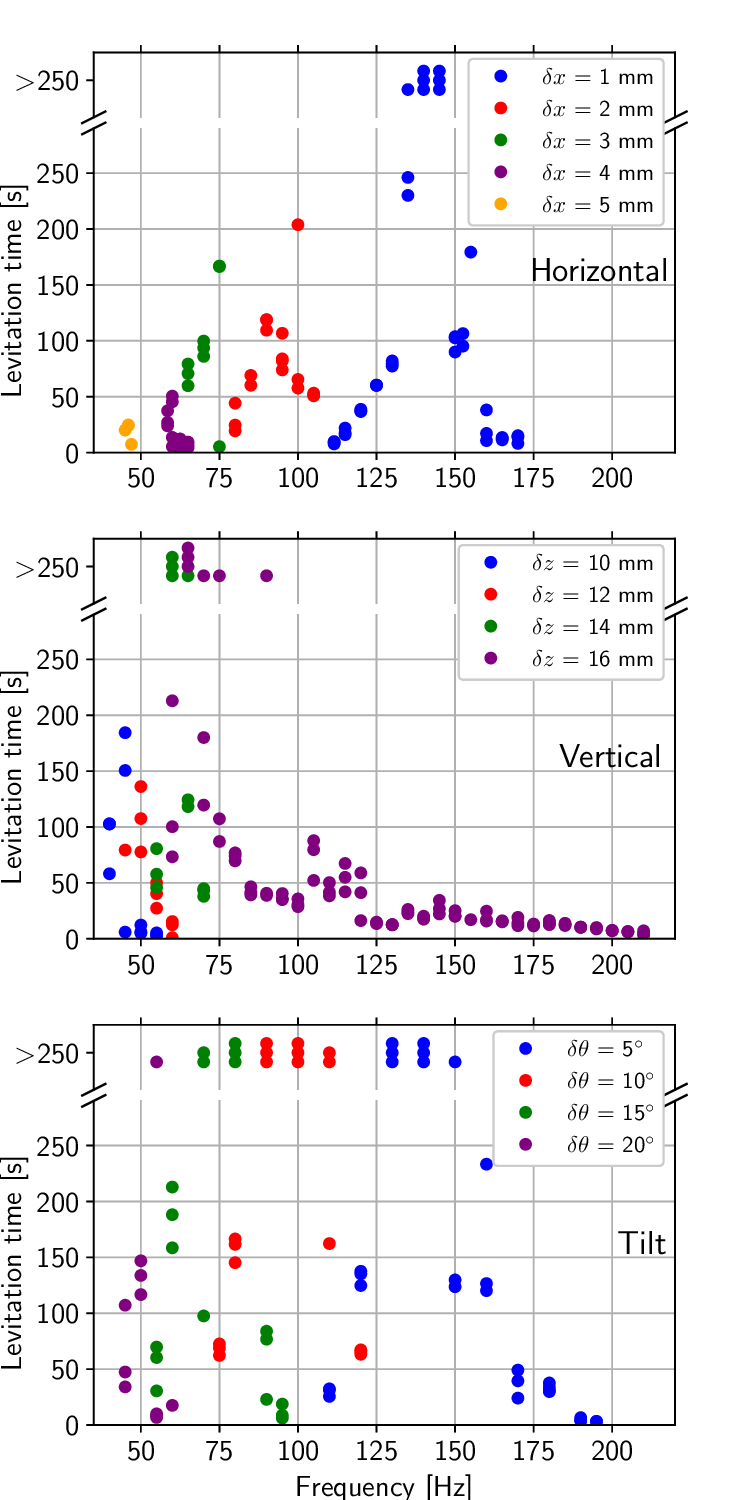}
\caption{The levitation time for the three different rotor configurations as function of rotor frequency, for the different geometric parameters. To limit the time-consumption of experiments, we stop the measurement if the levitation time is more than 250 s. Each experiment was repeated three times - repeated measurements of more than 250 s are displaced vertically, so all datapoints are visible.}
\label{Fig.Levitation_time} 
\end{center}
\end{figure}

In our previous study \cite{Hermansen_2023}, infinite levitation time was only achieved with an aluminium block near the floater, as the process of inducing eddy currents leads to a rotational damping that suppresses instabilities. However, with a static on-axis field applied by the permanent magnet configuration in the rotor, we here show that stable levitation is possible without added damping. Shown in Fig. \ref{Fig.Levitation_time} is the levitation time as function of frequency for the different rotor configurations. As can be seen from the figure, the lower the levitation frequency, the more unstable the levitation becomes in general. However, if even a small, static on-axis field is provided, the levitation can in multiple configurations be made to extend to very long times. To demonstrate that the levitation is stable, an experiment with a vertical displacement of $\delta{}z = 14$ mm at a frequency of 60 Hz was observed to levitate for 24 hours before the experiment was stopped. The behavior of levitation time as function of frequency in general follows the trend described in Ref. \cite{Hermansen_2023}, namely a linear increase on the low frequency side of the peak and a more rapid decline on the high frequency side of the peak. However, there are cases e.g. vertical displacement for $\delta{}z = 16$ mm where the trend is not reproduced.

In Fig. 4 in the supplementary material, we show three plots of the recorded levitation distance as function of time for the horizontal displacement with $\delta{}x=1$ mm. These exactly show a behavior of a linear fall rate at low frequencies, a stable levitation at selected frequencies and a highly oscillating levitation at large frequencies. We note that the simulation model always results in stable levitation, if this is possible.

In conclusion, we have shown that increasing the on-axis magnetic field in the Ucar effect increases the levitation distance and can lower the frequency needed to obtain levitation by a factor of 3. 
We showed this for three different rotor configurations, with three distinct magnetic fields. The experimental results were consistent with numerical simulations where Newtons equations of motion were solved and qualitatively agreed with simplified analytical expressions for the on-axis force. 

The lowering of rotation speed is important because it enables tuning of the Ucar effect and increases the range of available equipment; for instance 50 Hz is within the operating range of a standard drill. Also the minimum rotor speed is inversely dependent on floater radius \cite{Hermansen_2023}, meaning that for the fine manipulation of small objects, producing high enough frequencies could otherwise be problematic. Additionally, it will be possible to lift very large objects at modest rotation rates.

In future work it would be pertinent to extend the model so the full range of dynamical instabilities can be reproduced numerically, and to include the stabilizing effect of environmental eddy currents. Subsequently, an analysis with dynamical systems theory should yield more complete stability criteria.

\section*{Data statement}
All data presented in this work are available from Ref. \cite{Data_2025}.

\bibliographystyle{unsrt}

\clearpage

\begin{strip}
    \huge{Supplementary material for the article ``Magnetic levitation at low rotation frequencies using an on-axis magnetic field''}
\end{strip}

\begin{strip}
\setcounter{section}{0}
\setcounter{equation}{0}

\section{Magnetic field components at different levitation distance}
The magnetic field provided by the rotor at the location of the levitating magnet has been calculated using the magnetostatic framework MagTense \cite{Bjoerk_2021_supp}, which can fully account for the demagnetization field resulting from the geometry of the individual magnets in the rotor \cite{Smith_2010_supp}.

In Fig. \ref{Fig.H_axis} we show the magnetic field of the rotor at several possible locations of the floating magnet, as indicated in the figure legend. As can be seen, all different rotor arrangements produce an on-axis magnetic field, $\mu_{0}H_z$, at the location of the floating magnet. For the horizontal displacement and tilt, changing the geometrical parameter also changes the off-axis magnetic field.

\renewcommand{\thefigure}{1}
\renewcommand{\figurename}{Figure Supp.}
\begin{center}
\includegraphics[width=.60\textwidth]{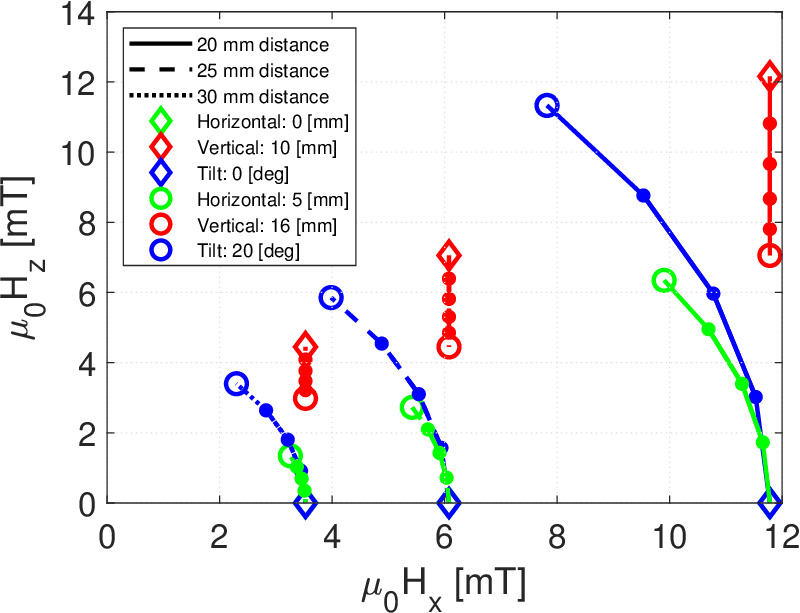}
\captionof{figure}{The two components of the magnetic field generated by the rotor at several possible locations of the floating magnet for the various rotor configurations investigated.}
\label{Fig.H_axis} 
\end{center}

\section{Off-axis oscillations of the floater magnet}
Our numerical simulations show that the radial restoring force present between the rotor magnet and the floater magnet results in a small circular motion of the floater magnet with a given radius, $a$. The stability along the axis of rotation can be explained by a competition between a repulsive and attractive dipolar force. Regarding the small circular, we prove below that magnetic interactions produce a net radial force in the plane perpendicular to the rotation axis, which precisely balances the centrifugal force from the circular motion; at least in our simulations.

We consider the situation where both the floater and rotor magnet are rotating counter clockwise at angular frequency $\omega$. Their magnetic moment and their displacement vector, $\bf{r}$, are given by
\begin{align}
    \bf{\hat{m}}_\text{r} = \begin{pmatrix}
    \cos(\omega_\text{r} t) \\ \sin(\omega_\text{r} t) \\ 0
    \end{pmatrix}
    ,\quad
    \bf{\hat{m}}_\text{f} = \begin{pmatrix}
    \sin(\theta_f)\cos(\omega_\text{r} t) \\ \sin(\theta_f)\sin(\omega_\text{r} t) \\ \cos(\theta_f)
    \end{pmatrix}
    ,\quad
    \bf{r} = \begin{pmatrix}
    a\cos(\omega_\text{r} t) \\ a\sin(\omega_\text{r} t) \\ -d
    \end{pmatrix}.
\end{align}
 Additionally, we assume the floater is moving in a circle around the rotation axis, also at frequency $\omega$, which corresponds to the small circular motion which we also term the side mode. This is illustrated in Fig. \ref{Fig.Side_mode}. The levitation distance along the rotation axis is $d = |\mathbf{r} \boldsymbol{\cdot} \mathbf{\hat{z}}|$.

\renewcommand{\thefigure}{2}
\begin{center}
\includegraphics[width=.30\textwidth]{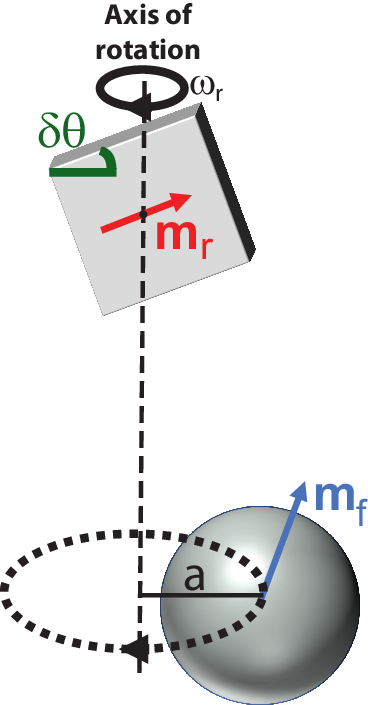}
\captionof{figure}{An illustration of the side mode motion of the floater, with radius $a$ of the circular motion. The size of $a$ with respect to the sizes of the magnets is greatly exaggerated. The floater has a diameter of 12.7 mm and $a$ is typically in the range 0.1-0.3 mm.}
\label{Fig.Side_mode} 
\end{center}

The dipole magnetic force is
\begin{align}
    \bf{F}_\text{dip} &= \frac{3\mu_0}{4\pi r^5} \left((\mathbf{m}_\text{f} \boldsymbol{\cdot} \mathbf{r}) \mathbf{m}_\text{r} + (\mathbf{m}_\text{r} \boldsymbol{\cdot} \mathbf{r})\mathbf{m}_\text{f} + (\mathbf{m}_\text{f} \boldsymbol{\cdot} \mathbf{m}_\text{r}) \mathbf{r} - \frac{5(\mathbf{m}_\text{f} \boldsymbol{\cdot} \mathbf{r})(\mathbf{m}_\text{r} \boldsymbol{\cdot} \mathbf{r})}{r^2} \mathbf{r} \right)
    \\
    &= - \mathcal{G} \frac{d}{r} \left(s_\theta \mathbf{\hat{z}} + c_\theta \mathbf{\hat{m}}_\text{r}\right) + \mathcal{G} \frac{a}{r} \left(3 s_\theta \mathbf{\hat{m}}_\text{r} + c_\theta \mathbf{\hat{z}} - 5 s_\theta \frac{a^2}{r^2} \mathbf{\hat{m}}_\text{r}\right) \quad \text{with} \quad \mathcal{G} = \frac{3\mu_0 m_\text{f} m_\text{r}}{4\pi r^4}.
\end{align}

Removing higher order terms, the transverse dipole force is given by
\begin{align}
    \mathbf{F}_\perp \approx - \mathcal{G} \mathbf{\hat{m}}_\text{r}.
\end{align}
As this force is always radial it results in steady circular motion. We can estimate the radius of the orbit, $a$, by
\begin{align}
\label{eq:sidemode}
    - \mu_\text{f} a \omega_r^2 \mathbf{\hat{m}}_\text{r} = \mathbf{F}_\perp \quad \xRightarrow{} \quad a = \frac{\mathcal{G}}{\mu_\text{f} \omega_r^2} = \frac{3\mu_0 m_\text{f} m_\text{r}}{4\pi d^4 \mu_\text{f} \omega_r^2},
\end{align}
where $\mu_\text{f}$ is the floaters mass. Note that the floater must be displaced in the $\mathbf{\hat{m}}_\text{r}$ direction for magnetic and centrifugal forces to cancel.
This expression for $a$ is valid for all of the experiments, but for the case of the vertical displacement (VD) experiments it is less trivial, so in section \ref{subsec:orbit_amplitude_with_vertical_displacement} we present a separate derivation for the VD case.

An illustration of the small circular motion is seen in Fig. \ref{Fig.Small_side_mode} which shows the position of the rotor, $x_r$ and $y_r$, and the floater, $x_f$ and $y_f$, coming from a simulation for a horizontal displacement of $\delta{}x=1$ mm and a frequency of 170 Hz. The rotor can be seen to have a radial amplitude of $1$ mm identical to the horizontal displacement. The floater can be seen to move in a circular motion. The amplitude of this motion as predicted from Eq. \eqref{eq:sidemode} is also shown and it is seen to agree very well with the simulation result.

\renewcommand{\thefigure}{3}
\begin{center}
\includegraphics[width=.60\textwidth]{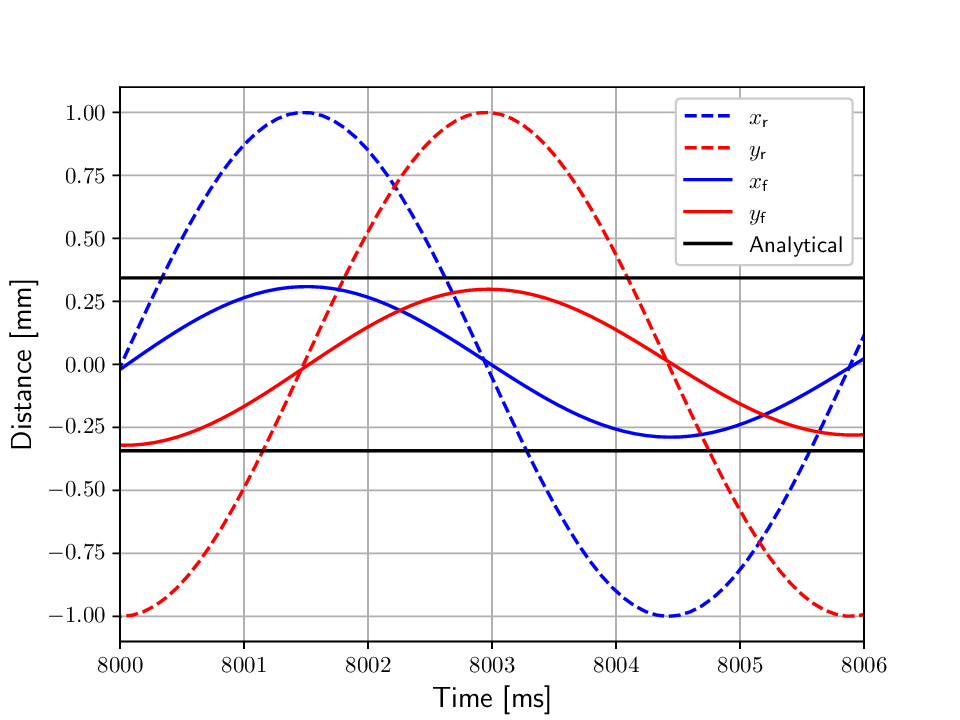}
\captionof{figure}{The position of the rotor,  $x_r$ and $y_r$, and the floater, $x_f$ and $y_f$, from the simulation for a horizontal displacement of $\delta{}x=1$ mm and a frequency of 170 Hz. Also shown in the amplitude of the circular motion of the floater, $a$, calculated from Eq. \eqref{eq:sidemode}.}
\label{Fig.Small_side_mode} 
\end{center}

\section{On-axis magnetic force}
In this section we derive the analytical expressions for the on-axis magnetic force, and also the analytical expressions for the levitation distance when possible. 

The magnetic dipole field provided by the rotor is given by
\begin{align}
   \mathbf{B}_\text{r} &= \frac{\mu_0}{4\pi r^3} \left(3[\hat{\mathbf{r}} \boldsymbol{\cdot} \mathbf{m}_\text{r}] \hat{\mathbf{r}} - \mathbf{m}_\text{r}\right) \label{Eq.Bfieldgeneral}
\end{align}
We denote the radial and $z$-components of the field as  $B_{\text{r}, \perp}$ and $ B_{\text{r}, z}$, respectively.

The dipole force on the floater is given by
\begin{align}
    \mathbf{F}_\text{dip} &= \frac{3\mu_0}{4\pi} \frac{1}{r^4}\big[(\mathbf{m}_\text{f} \mathbf{\cdot} \hat{\mathbf{r}}) \mathbf{m}_\text{r} + (\mathbf{m}_\text{r} \mathbf{\cdot} \hat{\mathbf{r}}) \mathbf{m}_\text{f} +(\mathbf{m}_\text{f} \mathbf{\cdot} \mathbf{m}_\text{r})\hat{\mathbf{r}}  -5(\mathbf{m}_\text{f}\mathbf{\cdot} \hat{\mathbf{r}})(\mathbf{m}_\text{r} \mathbf{\cdot} \hat{\mathbf{r}}) \hat{\mathbf{r}}\big] \label{eq:F_dip2}
\end{align}

We have previously, by solving Newtons equation of motion for the floater, found the polar angle of the floater, $\theta_\text{f}$, as \cite{Hermansen_2023_supp}
\begin{align}
    \theta_\text{f} = \frac{m_\text{f} B_{\text{r}, \perp}}{I_\text{f} \omega_\text{r}^2 - m_\text{f} B_{\text{r}, z}} \label{eq:theta&phi}
\end{align}
where $\omega_\text{r}$ is the angular velocity of the rotor and
$I_\text{f}$ is the moment of inertia of the floater given by 
\begin{align}
    I_\text{f}=\frac{8}{15}\pi \rho_\text{f} R_\text{f}^5
\end{align}

In the following we will derive the on-axis force for each of the rotor configurations studied.

\subsection{Horizontal displacement configuration} 
In this configuration the rotor magnet is displaced $\delta{}x$ from the rotation axis. We include the base side-mode derived in equation \eqref{eq:sidemode}, assume a small polar angle $\theta_\text{f} \ll 1$, and assume the levitation distance is much larger than both radial displacement and side mode radius, i.e. $d \gg \delta{}x, a$. The displacement vector is then
\begin{align}
    \mathbf{r} =\begin{pmatrix}
        -\delta{}x+a \\ 0 \\ -d
    \end{pmatrix}, 
\end{align}
with $r \approx d$ so $\mathbf{\hat{r}} \approx \mathbf{r}/d$.
Since our system is in steady state consider without loss of generality the point when the magnetization of the floater is in the $xz$-plane:
\begin{align}
    \mathbf{m}_\text{f} = m_\text{f} \begin{pmatrix}
        \sin \theta_\text{f} \\ 0 \\ -\cos\theta_\text{f}
    \end{pmatrix}  \quad \text{and} \quad \mathbf{m}_\text{r} = m_\text{r} \begin{pmatrix}
        1 \\ 0 \\ 0
    \end{pmatrix} 
\end{align}
Now we insert this into the $\mathbf{B}$-field and the dipole force expression which yields

\begin{align}
    \mathbf{F}_\text{hor} &= \mathcal{G}\Biggr[\left(\frac{a-\delta{}x}{d}\sin\theta_\text{f}-\cos\theta_\text{f}\right)  \begin{pmatrix}         1 \\ 0 \\ 0     \end{pmatrix} +  \frac{a-\delta{}x}{d}  \begin{pmatrix}        \sin\theta_\text{f} \\ 0 \\ \cos\theta_\text{f}     \end{pmatrix}
    \notag\\
    &+\sin \theta_\text{f} \begin{pmatrix}         \frac{a-\delta{}x}{d} \\ 0 \\ -1     \end{pmatrix}  -5\left(\frac{a-\delta{}x}{d}\sin \theta_\text{f}-\cos \theta_f \right)\frac{a-\delta{}x}{d} \begin{pmatrix}         \frac{a-\delta{}x}{d} \\ 0 \\ -1     \end{pmatrix}\Biggr].  \label{eq:F_hor}
\end{align}
In the $z$-direction we thus get
\begin{align}
    F_\text{z,hor} &= \mathcal{G} \Biggr[ 
    \left(5\frac{(\delta{}x-a)^2}{d^2}-1 \right)\sin \theta_\text{f} + 4\frac{\delta{}x-a}{d} \cos \theta_f \Biggr].    \label{eq:F_hor_z}
\end{align}
Invoking $\theta_\text{f}\ll 1$ and using $\delta x - a \ll d$ the vertical force in the horizontal configuration is to a good approximation

\begin{align}
    \label{eq:HD_forceEq}
    F_\text{z,hor} &= \mathcal{G}\left( 4\frac{\delta{}x-a}{d} 
     -\theta_\text{f}   \right). 
\end{align}

From Eq. \eqref{Eq.Bfieldgeneral}, one may show that in the present configuration
\begin{align}
    B_{\text{r}, \perp}=\frac{\mu_0 m_\text{r}}{4\pi d^3} \quad \text{and} \quad B_{\text{r}, z} = \frac{\delta{}x}{d} B_{\text{r}, \perp} \label{eq:B_field_horizontal}
\end{align}
to leading order in $\delta x / d$. Thus $B_{\text{r},z} \ll B_{\text{r},\perp}$ so the assumption $\theta_\text{f} \ll 1$ requires $I_\text{f} \omega_\text{r}^2 \gg m_\text{f} B_{\text{r},\perp} \gg m_\text{f} B_{\text{r},z}$ (cf. Eq. \eqref{eq:theta&phi}). It follows that
\begin{align}
    \theta_\text{f} \approx \frac{m_\text{f} B_{\text{r},\perp}}{I_\text{f} \omega_\text{r}^2} = \frac{G}{d^3 I_\text{f}\omega_\text{f}^2} \quad \text{where} \quad G = \frac{\mu_0 m_\text{f} m_\text{r}}{4\pi}.   \label{Eq.theta_f_approx}
\end{align}
Inserting this into Eq. \eqref{eq:HD_forceEq} and recalling that $a$ is a function of $d$ via Eq. \eqref{eq:sidemode}, we can obtain the steady-state levitation distance by solving $F_\text{z,hor} = 0$ for $d$. This yields a second order polynomial in $d^2$ with the solution
\begin{align}
    d_\text{hor}^2 = \frac{G}{8\delta x I_\text{f} \omega_\text{f}^2} + \sqrt{\left(\frac{G}{8\delta x I_\text{f}\omega_\text{f}^2}\right)^2 + \frac{3 G}{\delta x \mu_\text{f} \omega_\text{r}^2}}. \label{Eq.d_hor}
\end{align}
Interestingly, every term depends on the ratio of $G$ (the magnetic interaction strength) to either $I_\text{f} \omega_\text{f}^2$ (the floaters angular momentum) or $\mu_\text{f} \omega_\text{r}^2$ (its linear momentum), and when $\delta x \xrightarrow{} 0$ (no displacement) the distance diverges.

\subsection{Vertical displacement configuration}
In the vertical displacement configuration, there are two magnets which generate the magnetic field of the rotor. The magnetic moment is denoted by $m_\text{r}^\text{H}$ for the horizontally oriented magnet and $m_\text{r}^\text{V}$ for the vertical magnet. In steady-state at exactly the point when the magnetization of the floater magnet is in the $xz$-plane these are given by
\begin{align}
    \mathbf{m}_\text{f} = m_\text{f} \begin{pmatrix}
        \sin \theta_\text{f} \\ 0 \\ \cos\theta_\text{f}
    \end{pmatrix}  \quad \text{,} \quad \mathbf{m}^\text{H}_\text{r} =m^\text{H}_\text{r} \begin{pmatrix} 
        1 \\ 0 \\ 0
    \end{pmatrix} \text{and} \quad \mathbf{m}^\text{V}_\text{r} =m^\text{V}_\text{r} \begin{pmatrix} 
        0 \\ 0 \\ 1
    \end{pmatrix} 
\end{align}

The vertical magnet is located $\delta{}z$ above the horizontal magnet. The displacement vectors from the center of the rotor magnets to the floater magnet are
\begin{align}
    \mathbf{r}^\text{H} =\begin{pmatrix}
        a \\ 0 \\ -d
    \end{pmatrix} \quad \text{and} \quad
    \mathbf{r}^\text{V} =\begin{pmatrix}
        a \\ 0 \\ -(d+\delta{}z)
    \end{pmatrix}
\end{align}
We assume $a \ll d$, so $r^\text{H} \approx d$ and $r^\text{V} \approx d+\delta z$.

The magnetic field of the rotor, $\mathbf{B}_\text{r}$, is given by 
\begin{align}
    \mathbf{B}_{\text{r}} &=
    \frac{\mu_0 m^\text{H}_\text{r}}{4\pi d^3} \left[3
        \frac{a}{d}  \begin{pmatrix}
        \frac{a}{d} \\ 0 \\ -1
    \end{pmatrix} - \begin{pmatrix} 
        1 \\ 0 \\ 0
    \end{pmatrix} \right] +
    \frac{\mu_0 m^\text{V}_\text{r}}{4\pi (d+\delta{}z)^3} \left[  -3  
       \begin{pmatrix}
        \frac{a}{d+\delta{}z} \\ 0 \\ -1
    \end{pmatrix} -\begin{pmatrix} 
        0 \\ 0 \\ 1
    \end{pmatrix} \right]
\end{align}

i.e. with off-axis and on-axis components given by
\begin{align}
    \label{eq:B_field_VD}
    B_{\text{r}, \perp} &= \frac{\mu_0 m^\text{H}_\text{r}}{4\pi d^3} \left(  3
        \frac{a^2}{d^2}  -1 \right) -    \frac{3\mu_0 m^\text{V}_\text{r} a}{4\pi (d+\delta{}z)^4} \nonumber\\
    B_{\text{r}, z} &= -\frac{3\mu_0 m^\text{H}_\text{r}a}{4\pi d^4}  +
    \frac{2\mu_0 m^\text{V}_\text{r}}{4\pi (d+\delta{}z)^3}
\end{align}

Inserting these into the equation for the polar angle, Eq. \ref{eq:theta&phi}, we get
\begin{align}
\label{eq:theta_assumption_VD}
    \theta_\text{f} &= \frac{m_\text{f} \frac{\mu_0 m^\text{H}_\text{r}}{4\pi d^3} \left(  3
        \frac{a^2}{d^2}  -1 \right) -    \frac{3\mu_0 m^\text{V}_\text{r} a}{4\pi (d+\delta{}z)^4}}{I_\text{f} \omega_\text{r}^2 - m_\text{f} \left( -\frac{3\mu_0 m^\text{H}_\text{r} a}{4\pi d^4}  +
    \frac{2\mu_0 m^\text{V}_\text{r}}{4\pi (d+\delta{}z)^3} \right)}
\end{align}

Since there are two rotor magnets we calculate the forces separately and add these up afterwards. For the horizontally magnetized rotor magnet, the force is the same as Eq. \eqref{eq:F_hor} when setting $\delta x = 0$, so the $z$-component is (cf. Eq. \eqref{eq:F_hor_z})
\begin{align}
\label{eq:VD_forcez_H}
    F^\text{H}_\text{z,ver}&= \mathcal{G} \Biggr[\left(5\frac{a^2}{d^2} - 1\right) \sin \theta_\text{f} - 4 \frac{a}{d} \cos \theta_\text{f}   \Biggr]
\end{align}

We now consider the vertically magnetized rotor magnet. The dipole force from this is
\begin{align}
   \mathbf{F}^\text{V}_\text{ver} =& \frac{3\mu_0 m_\text{f}m_\text{r}}{4\pi (d+\delta{}z)^4} \Biggr[\left(  \frac{a}{d+\delta{}z} \sin \theta_\text{f} -\cos \theta_\text{f} \right) \begin{pmatrix}         0 \\ 0 \\ 1      \end{pmatrix} -    \begin{pmatrix}
        \sin \theta_\text{f} \\ 0 \\ \cos\theta_\text{f}
    \end{pmatrix} 
    \notag\\
     &+ \cos \theta_\text{f} \begin{pmatrix}         \frac{a}{d+\delta{}z} \\ 0 \\ -1     \end{pmatrix}   +5 \left(  \frac{a}{d+\delta{}z} \sin \theta_f -\cos \theta_f \right) \begin{pmatrix}         \frac{a}{d+\delta{}z} \\ 0 \\ -1     \end{pmatrix} \Biggr]
\end{align}

The force in the $z$-direction is thus 
\begin{align}
\label{eq:VD_forcez_V}
    F^V_\text{z,ver}&= \frac{3\mu_0 m_\text{f}m_\text{r}}{4\pi (d+\delta{}z)^4} \Biggr(-4 \frac{a}{d+\delta{}z} \sin \theta_\text{f}   + 2\cos \theta_\text{f} \Biggr)
\end{align}

Combining the two forces from Eqs. \eqref{eq:VD_forcez_H} and \eqref{eq:VD_forcez_V}, and assuming $\theta_\text{f}$ is small the force in the $z$-direction can be approximated as
\begin{align}
    F_\text{z,ver}&= \frac{3\mu_0 m_\text{f}m_\text{r}}{4\pi } \Biggr( \frac{2}{(d+\delta{}z)^4} +  4 \frac{\delta{}x-a}{d^5} - \frac{\theta_\text{f}}{d^4}   \Biggr)
\end{align}
Unfortunately, even with the $\theta_\text{f} \ll 1$ assumption, the above expression cannot be solved to find a closed-form expression for the levitation distance. \newline

\subsubsection{The amplitude of the circular motion for the vertical displacement configuration \label{subsec:orbit_amplitude_with_vertical_displacement}}
In order to find the radius of the small circular motion, $a$, for the two-magnet configuration present in the vertical displacement case, we look at the dipole forces perpendicular to the rotation axis, which are given by
\begin{align}
    F^\text{H}_{\perp} &= \frac{3\mu_0 m_\text{f} m_\text{r}}{4\pi d^4}\Biggr[ 3\frac{a}{d} \sin \theta_\text{f} -\cos\theta_\text{f}  
    -5 \frac{a^2}{d^2}\left(\frac{a}{d} \sin \theta_\text{f} - \cos \theta_\text{f} \right)       \Biggr]
\end{align}
and
\begin{align}
    F^\text{V}_{\perp} =& \frac{3\mu_0 m_\text{f}m_\text{r}}{4\pi (d+\delta{}z)^4} \Biggr[-\sin \theta_\text{f} + \frac{a}{d + \delta z}\cos \theta_\text{f} + 5\left(  \frac{a}{d+\delta{}z} \sin \theta_\text{f} - \cos \theta_\text{f} \right) \frac{a}{d+\delta{}z}  \Biggr]
\end{align}
Combining these we find the force:
\begin{align}
    F_{\perp} &= \frac{3\mu_0 m_\text{f}m_\text{r}}{4\pi} \Biggr[ 3 \frac{a}{d^5} \sin \theta_\text{f} -\frac{\cos \theta_\text{f}}{d^4}   
    + 5\frac{a^2}{d^6} \left(  -\frac{a}{d} \sin \theta_\text{f} + \cos \theta_\text{f} \right) \nonumber  \\ 
    &-\frac{\sin \theta_\text{f}}{(d+\delta{}z)^4} + \frac{a}{(d+\delta{}z)^5} \cos \theta_\text{f} 
    +\frac{5a}{(d+\delta{}z)^5} \left(  \frac{a}{d+\delta{}z} \sin \theta_\text{f} - \cos \theta_\text{f} \right)       \Biggr]
\end{align}
To leading order in $a/d$ this simplifies to
\begin{align}
    F_{\perp} &= \mathcal{G} \Biggr( \cos \theta_\text{f}   
    +\frac{d^4}{(d+\delta{}z)^4} \sin \theta_\text{f} \Biggr)
\end{align}
Inserting in Eq. \eqref{eq:sidemode} and solving for $a$:
\begin{align*}
      a = \frac{\mathcal{G}}{\mu_\text{f} \omega_r^2 } \Biggr(\cos \theta_\text{f}   
    +\frac{d^4}{(d+\delta{}z)^4} \sin \theta_\text{f} \Biggr) \approx  \frac{\mathcal{G}}{ \mu_\text{f} \omega_r^2} 
\end{align*}
The approximated expression where we used $\theta_\text{f} \ll 1$ is the same as Eq. \eqref{eq:sidemode}.

\subsection{Tilt configuration}
In the tilt configuration, the rotor magnet is tilted $\delta\theta$ from horizontal. The displacement vector is
\begin{align}
    \mathbf{r} = \begin{pmatrix}
        a \\ 0 \\ -d
    \end{pmatrix} 
\end{align}
As above, let $d \gg a$ so that $r \approx d$.
 
Again, considering the steady state moment when the magnetization of the floater magnet is in the $xz$-plane, the magnetic moments are given by
\begin{align}
    \mathbf{m}_\text{f} = m_\text{f} \begin{pmatrix}
        \sin \theta_\text{f} \\ 0 \\ \cos\theta_\text{f}
    \end{pmatrix}  \quad \text{and} \quad \mathbf{m}_\text{r} = m_\text{r} \begin{pmatrix}
        \cos{\delta\theta} \\ 0 \\ \sin{\delta\theta} 
    \end{pmatrix} 
\end{align}

Inserting these into Eqs. \eqref{Eq.Bfieldgeneral} and \eqref{eq:F_dip2} for the $\mathbf{B}$ field and dipole force, we get the force
\begin{align}
    &\mathbf{F}_\text{tilt} = \mathcal{G} \Bigg[ \left(\frac{a}{d}\sin \theta_\text{f} -\cos \theta_\text{f}\right)\begin{pmatrix}         \cos \delta\theta  \\ 0 \\ \sin \delta\theta      \end{pmatrix} + \left(\frac{a}{d} \cos \delta\theta -\sin \delta\theta \right)   \begin{pmatrix}         \sin \theta_\text{f} \\ 0 \\ \cos \theta_\text{f}     \end{pmatrix} 
    \quad \nonumber\\
    &+ \sin(\theta_\text{f} + \delta \theta) \begin{pmatrix}         a/d \\ 0 \\ -1     \end{pmatrix} -5\left( \frac{a}{d} \sin \theta_\text{f} -\cos\theta_\text{f}\right)\left(\frac{a}{d}\cos \delta\theta - \sin \delta\theta\right)   \begin{pmatrix}         a/d \\ 0 \\ -1     \end{pmatrix} \Bigg] 
\end{align}

In the $z$-direction, this amounts to
\begin{align}
    F_\text{z,tilt} &= \mathcal{G} \Bigg[2 \cos \theta_\text{f} \sin \delta \theta - 4 \frac{a}{d} \cos(\theta_\text{f} - \delta \theta) + \left(5 \frac{a^2}{d^2} - 1\right) \sin \theta_\text{f} \cos \delta \theta \Bigg] 
\end{align}

Assuming $\theta_\text{f}, \delta \theta \ll 1$ and only neglecting the term proportional to $\frac{a^2}{d^2}$, we get
\begin{align}
    F_\text{z,tilt} &= \mathcal{G} \left( 2\delta\theta-\theta_\text{f} - 4\frac{a}{d} \right).
\end{align}
We also obtain the same $\theta_\text{f}$ approximation as in Eq. \eqref{Eq.theta_f_approx}, hence
\begin{align*}
    F_{z,\text{tilt}} = \mathcal{G} \left(2\delta \theta - 4\frac{a}{d} - \frac{G}{d^3 I_\text{f}\omega_\text{f}^2} \right)
\end{align*}

If we further neglect the $a/d$ term, we can solve the equation $F_{z,\text{tilt}} = 0$ for levitation distance. The result is
\begin{align}
    d_\text{tilt}=\left(\frac{G}{2I_\text{f}\omega_\text{f}^2 \delta \theta}\right)^\frac{1}{3},    \label{Eq.d_tilt}
\end{align}
in agreement with eq. 12 of Ref. \cite{le_lay_magnetic_2024_supp}. The last simplification is poorly justified, which may explain why Eq. \eqref{Eq.d_hor} matches the simulated and experimental data better than Eq. \eqref{Eq.d_tilt}.
We note that $d_\text{tilt}$ depends on the ratio of the interaction strength $G$ and the angular momentum $I_\text{f} \omega_\text{f}^2$, and it diverges without the tilt i.e.\ when $\delta \theta \xrightarrow{} 0$.

\clearpage

\section{Levitation trajectory}
For all experiments performed, the levitation trajectory were recorded, i.e. the levitation distance as function of time for the floater. These are too numerous to show, but in general they fall in three categories:
\begin{itemize}
    \item Linear fall rate
    \item Stable levitation
    \item Unstable levitation
\end{itemize}

In Fig. \ref{Fig.Trajectories} we show examples of these three types of trajectories for the case of $\delta{}x = 1$ mm for the horizontal displacement experiment. For a low frequency, $f=115$ Hz, the floater has a linear fall rate until it gets so far away from the rotor that levitation ends. For a frequency of $f=140$ Hz, the levitation is stable. For $f=150$ Hz the levitation is unstable and the floater oscillates significantly until it gets so far away from the rotor that levitation ends.

\renewcommand{\thefigure}{4}
\begin{center}
\includegraphics[width=.60\textwidth]{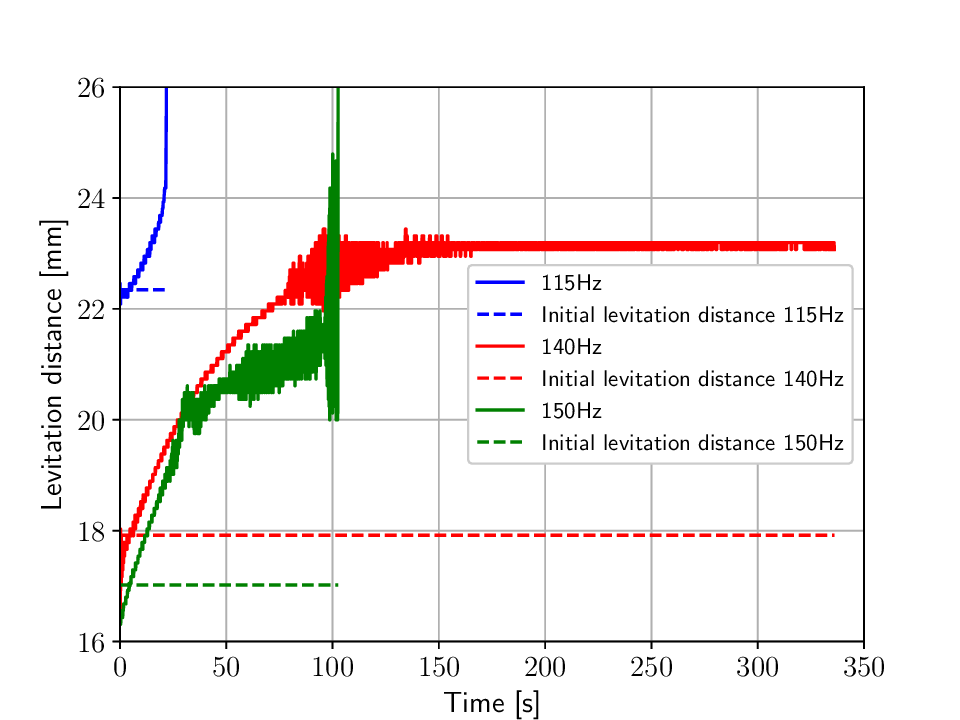}
\captionof{figure}{Three trajectories, i.e. levitation distance as function of time, for different frequencies for the case of $\delta{}x = 1$ mm for the horizontal displacement experiments.}
\label{Fig.Trajectories} 
\end{center}

\end{strip}

\bibliographystyle{unsrt}

\end{document}